\documentclass[fleqn,usenatbib]{mnras}

\usepackage{newtxtext,newtxmath}

\usepackage[T1]{fontenc}
\usepackage{ae,aecompl}

\usepackage{physics}
\usepackage{cleveref}

\usepackage{graphicx}	
\usepackage{amsmath}	
\usepackage{amssymb}	
\usepackage[normalem]{ulem}

\graphicspath{{./figures/}}

\title[Radius valley: a by-product of planet formation]{Sculpting the Valley in the Radius Distribution of Small Exoplanets as a by-product of Planet Formation: The Core-Powered Mass-Loss Mechanism}

\author[A. Gupta and H.E. Schlichting]{
Akash Gupta$^{1}$\thanks{E-mail: akashgpt@ucla.edu} and
Hilke E. Schlichting$^{1,2}$
\\
$^{1}$Department of Earth, Planetary, and Space Sciences, University of California, Los Angeles, CA 90095, USA\\
$^{2}$Department of Earth, Atmospheric and Planetary Sciences, Massachusetts Institute of Technology, MA 02139, USA
}

\date{Accepted XXX. Received YYY; in original form ZZZ}

\pubyear{2019}

\begin{document}
\label{firstpage}
\pagerange{\pageref{firstpage}--\pageref{lastpage}}
\maketitle

\begin{abstract}
Recent observations revealed a bimodal radius distribution of small, short-period exoplanets with a paucity in their occurrence, a radius `valley', around $1.5-2.0$ $R_\oplus$. In this work, we investigate the effect of a planet's own cooling luminosity on its thermal evolution and atmospheric mass-loss (core-powered mass-loss) and determine its observational consequences for the radius distribution of small, close-in exoplanets. Using simple analytical descriptions and numerical simulations, we demonstrate that planetary evolution based on the core-powered mass-loss mechanism alone (i.e., without any photoevaporation) can produce the observed valley in the radius distribution. Our results match the valley's location, shape and slope in planet radius-orbital period parameter space, and the relative magnitudes of the planet occurrence rate above and below the valley. We find that the slope of the valley is, to first order, dictated by the atmospheric mass-loss timescale at the Bondi radius and given by $\text{d log} R_p/ \text{d log} P \simeq 1/(3(1-\beta))$ {which evaluates to} $  -0.11$ {for $ \beta \simeq 4$}, where $M_c/M_\oplus = (R_c/R_{\oplus})^{\beta} (\rho_{c*}/\rho_{\oplus})^{\beta/3}$ is the mass-radius relation of the core. {This choice for $\beta$} yields good agreement with observations {and attests} to the significance of internal compression for {massive} planetary cores. We further find that the location of the valley scales as $\rho_{c*}^{-4/9}$ and that the observed planet population must have predominantly rocky cores with typical water-ice fractions of less than $\sim 20\%$. Furthermore, we show that the relative magnitude of the planet occurrence rate above and below the valley is sensitive to the details of the planet-mass distribution but that the location of the valley is not. 
\end{abstract}


\section{Introduction} \label{sec:intro}

NASA's \textit{Kepler} mission has unveiled a wealth of new planetary systems \citep[e.g.,][]{BK10}. These systems offer new insights into the process of planet formation and evolution. One of \textit{Kepler}'s key findings is that the most common planets in our galaxy{, observed to date,} are between 1 and 4 $R_\oplus$, i.e., larger than Earth but smaller than Neptune \citep{FT13,PM13}. Further observations revealed a transition in average densities at planet sizes $\sim 1.5 \;R_\oplus$ \citep{marcy2014b, rogers2015a}, with smaller planets having densities consistent with rocky compositions while larger planets having lower densities indicating significant H/He envelopes. {In addition, \citet{owen2013a} noticed a bimodal distribution of observed planet radii.} Since then, refined measurements have provided strong observational evidence for the sparseness of short-period planets in the size range of $\sim 1.5 - 2.0 \;R_\oplus$ relative to the smaller and larger planets, yielding a valley in the small exoplanet radius distribution \citep[e.g.,][]{fulton2017a,fulton2018a}. For example, the California-\textit{Kepler} Survey reported measurements from a large sample of 2025 planets, detecting a factor of $\sim 2$ deficit in the relative occurrence of planets with sizes  $\sim 1.5 - 2.0 \;R_\oplus$ \citep{fulton2017a}. Studies suggest that this valley likely marks the transition from the smaller rocky planets: `super-Earths', to planets with significant H/He envelopes typically containing a few percent of the planet's total mass: `sub-Neptunes' \citep[e.g.,][]{LF13,owen2013a,LF14,rogers2015a,ginzburg2016a}. Furthermore, the location of this valley is observed to decrease to smaller planet radii, $R_p$, with increasing orbital period, $P$. In a recent study involving asteroseismology-based high precision stellar parameter measurements for a sample of 117 planets, a slope $\text{d log} R_p/ \text{d log} P = -0.09^{+0.02}_{-0.04}$ was reported for the radius valley { by \citet{eylen2018a}. A similar value for the slope of $-0.11^{+0.03}_{-0.03}$ was reported by \citet{MCG19}}.

{The observed valley in the exoplanet radius distribution has been attributed to photoevaporation of H/He atmospheres by high energy stellar radiation} \citep[e.g.,][]{owen2013a,LF13}. Recent work showed that thermal evolution models with photoevaporation can reproduce the observed radius distribution \citep[e.g.,][]{owen2017a,eylen2018a}.

However, photoevaporation by high energy photons is not the only proposed mechanism for shaping the radius valley. \citet{ginzburg2018a} demonstrated that the core-powered mass-loss mechanism  \citep{ginzburg2016a} can produce the exoplanet radius distribution, even without photoevaporation, solely as a by-product of the planet formation process itself. In the core-powered mass-loss mechanism, it is the luminosity of the cooling planetary cores that provide the energy for atmospheric loss.
{ The assembly of planetary cores results in large core temperatures as gravitational binding energy is converted into heat. Furthermore, if this assembly takes place in the presence of a gas disk, planetary cores not only accrete H/He atmospheres, but they are also prevented from cooling significantly since the optically thick H/He envelopes act like thermal blankets regulating the heat loss from both the core and envelope at the radiative-convective boundary \citep[e.g.][]{lee2015a,ginzburg2016a}. As a result, the temperature of super-Earth and sub-Neptune cores is dictated by the maximum temperatures that still allows for the accretion of H/He envelopes onto the core. This temperature is approximately given by $T_c \sim G M_c \mu/k_B R_c$, where $\mu$ is the mean molecular weight of the atmosphere, $k_B$ the Boltzmann constant, $G$ the gravitational constant and $M_c$ and $R_c$ are the mass and radius of the planetary core, respectively. These core temperatures evaluate to about 10,000-100,000~K for planets with masses between Earth and Neptune and these cores and envelopes take Gyrs to cool \citep{ginzburg2016a}.} 

In this paper, we extend the results from \citet{ginzburg2018a} and show that the core-powered mass-loss mechanism can produce the valley's location, shape and slope in planet radius-orbital period parameter space, and the relative magnitudes of the planet occurrence rate above and below the valley. We further use it to constrain the planet's core composition and the mass-radius relation of the core. This paper is structured as follows: \Cref{sec:model} is divided into several parts. In the first, we describe our planetary model, its structure and relevant equations. We then outline the core-powered mass-loss mechanism and define the parameters of the exoplanet population used in our numerical evolution calculations. We discuss our results in \Cref{sec:results}, which includes a comparison with observations and investigations into how our results depend on the physical parameters of the planet population. Our conclusions are summarized in \Cref{sec:conclusion}.

\section{Planet Structure and Evolution} \label{sec:model}
In this section, we describe our model for the structure of the core and envelope of close-in super-Earths and sub-Neptunes and their evolution due to core-powered mass-loss after the dispersal of the gas disk. For a detailed review of this mechanism, the reader is referred to \citet{ginzburg2016a} and \citet{ginzburg2018a}.

\begin{figure*}
\centering
\includegraphics[width=1\textwidth,trim=0 0 0 0,clip]{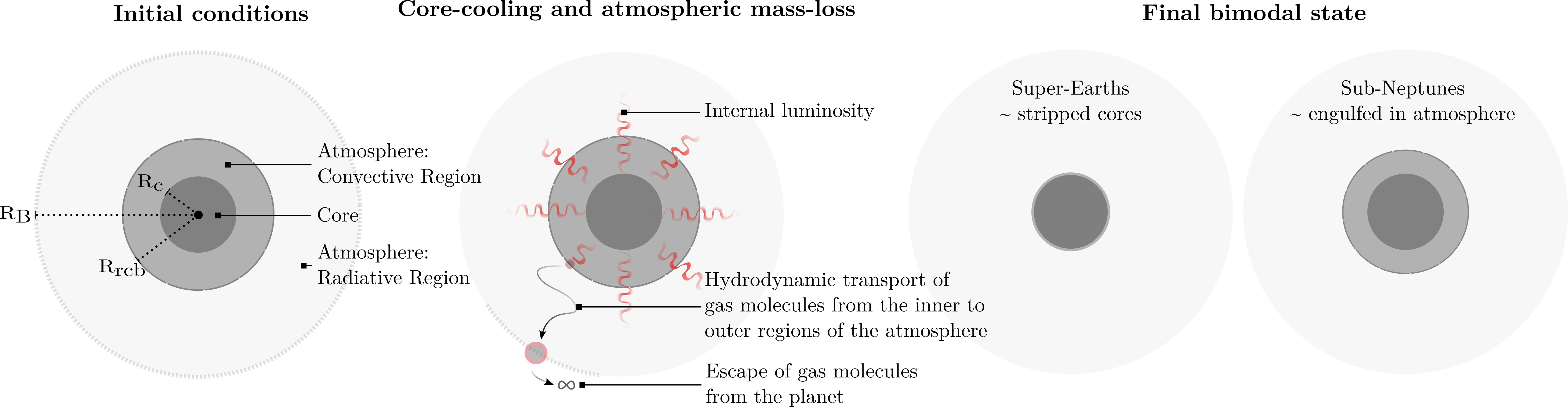}
\caption{Schematic of the main stages in the evolution of a planet due to the core-powered mass-loss mechanism. \textit{Left panel}: The  primary components of the planet structure: core (dark gray), and atmospheric convective (gray) and radiative (light gray) regions. As shown, the convective region of the atmosphere extends from the core to the radiative-convective boundary, $R_{rcb}$, which is comparable to a few core radii, $R_c$, at the end of the disk dispersal phase (our initial condition), and the radiative region extends from the $R_{rcb}$ to the Bondi radius, $R_B$. \textit{Middle panel}: Illustration of the thermal evolution and atmospheric mass-loss at the Bondi radius. \textit{Right panel}: The two end-member states at the end of 3 Gyrs. of evolution: (i) super-Earths, stripped rocky cores found below the valley, and (ii) sub-Neptunes, engulfed in H/He atmospheres and located above the valley.}
\label{fig:evo_model}
\end{figure*}

\subsection{Planet Structure}\label{sec:planet_structure}
We assume a planet of radius $R_p$ and mass $M_p$ with a dense core surrounded by a gaseous atmosphere, with most its mass in the core such that $M_c \sim M_p$. We note here that by core we mean the non-gaseous part of the planet and such that this includes both the iron core and silicate mantle of an Earth analog.
We assume, accounting for gravitational compression, that the mass-radius relationship for the core is given by ${M_c}/{M_\oplus} = (R_c/R_\oplus)^4(\rho_{c*}/\rho_\oplus)^{4/3}$ , where $\rho_{c*}$ is the {density of the core scaled to an Earth mass}, $R_c$ is the radius of the core and $\oplus$ refers to the corresponding Earth values \citep{valencia2006a, fortney2007a}. For pure water/ice, silicate and iron cores, we assume $\rho_{c*}$ to be 1.3 g cm$^{-3}$, 4 g cm$^{-3}$ and 11 g cm$^{-3}$, respectively {\citep{fortney2007a}}.

We assume that all cores are initially surrounded by H/He envelopes of mass $M_{atm}$ and define the atmosphere to core mass fraction as $f = M_{atm}/M_c$. As has been shown in previous works \citep[e.g.,][]{PY14,lee2015a,IS15}, the structure of the atmosphere is, to first order, well described by an inner convective region that contains most of the atmospheric mass and an outer radiative, nearly-isothermal, region of negligible mass. The transition between these two regions occurs at the radiative-convective boundary which we denote as $R_{rcb}$. We treat the $R_{rcb}$ as the planet's effective radius, i.e., $R_p \simeq R_{rcb}$. This is a good approximation as the density profile changes sharply at the $R_{rcb}$. We model the atmosphere as an ideal gas.

The dispersal of the protoplanetary disk causes a loss of pressure support on the outer edge of the envelope, causing atmospheric mass-loss powered by the luminosity of the cooling inner regions of the atmosphere \citep{owen2016a,ginzburg2016a}. As a result, the envelopes of close-in planets rapidly shrink (roughly on the timescale on which the disk disperses) to thicknesses $\Delta R \simeq R_c$, where $\Delta R = R_{rcb}-R_c$ is the thickness of the envelope measured from the core's surface. Since we are interested in the evolution of the planets after disk dispersal, we assume $\Delta R \simeq R_c$ as initial condition for the thickness of the planetary envelopes \citep[see also][]{owen2017a}.

The atmospheric mass can be obtained by integrating the density profile over the convective region, which yields
\begin{equation}\label{eq:M_atm}
M_{atm} = \frac{\gamma -1}{\gamma} 4\pi R_c^2 \rho_{rcb} {\Delta R} \left( \frac{R_B^\prime {\Delta R} }{R_c^2}\right)^{1/(\gamma -1)},
\end{equation}
where $\gamma$ is the adiabatic index of the atmosphere, $\rho_{rcb}$ is the density of the atmosphere at $R_{rcb}$ and $R_B^\prime$ is the modified Bondi radius \citep{ginzburg2016a}, such that
\begin{equation}
R_B^{\prime} \equiv \frac{\gamma - 1}{\gamma} \frac{G M_c \mu}{k_B T_{rcb}}
\end{equation}
where $\mu$ is the molecular mass of the atmosphere, $k_B$ is Boltzmann constant, $G$ is the gravitational constant and $T_{rcb} \sim T_{eq}$ is the temperature at the $R_{rcb}$, and $T_{eq}$ the equilibrium temperature for a given distance from the host star. Throughout this study, we assume Sun-like host stars.

The temperature at the base of the envelope is 
\begin{equation}
T_c \simeq \frac{\gamma - 1}{\gamma} \frac{1}{k_B} \frac{G M_c \mu}{R_c^2} \Delta R,
\end{equation}
which is valid for $R_c/R_B' \lesssim \Delta R/R_c \lesssim 1$ \citep{ginzburg2018a}. We ignore the ultra-thin regime, $\Delta R/R_c \lesssim R_c/R_B' \sim 0.1$, for which $T_c \sim T_{eq}$ as this regime cannot yet be detected in the exoplanet radius distribution. We also ignore any additional heat generated by radioactive decay in the core. Including this would delay the transition to the ultra-thin regime further, and it may lead to additional atmospheric mass-loss for planets whose loss is not limited by the cooling time or age of the system.

We model the core, to first order, as incompressible, molten and fully convective such that its temperature is close to isothermal and given by the temperature at the bottom of the convective region, $T_c$. We assume that the core-envelope interface is well coupled such that the core temperature always matches that of the base of the adiabatic atmosphere.

As a result, the thermal and gravitational energy available for cooling is
\begin{equation}\label{eq:E_cool}
E_{cool} \simeq g \Delta R \left( \frac{\gamma}{2\gamma -1}M_{atm} + \frac{1}{\gamma}\frac{\gamma-1}{\gamma_c -1} \frac{\mu}{\mu_c} M_c \right),
\end{equation}
where $\gamma_c$ and $\mu_c$ are the adiabatic index and molecular mass of the core, respectively, and $g=GM_c/R_c^2$ is the surface gravity of the planet. The first and second term on the right-hand side correspond to the atmosphere's energy and core's thermal energy, respectively.

\subsection{Evolution Model}\label{sec:CPM}

We start our evolution models right after the disk dispersal phase.
To distinguish our results from any atmospheric mass-loss due to photoevaporation, we only consider the planet's evolution due to its own cooling luminosity and its subsequent mass-loss. 

As shown above, the core temperatures are, as a result from formation, of the order of $10^4-10^5$ K. Since the core-envelope interface is well coupled, the cooling of both the core and envelope is dictated by {radiative} diffusion through the radiative-convective boundary. This implies that the planet cools at a rate
\begin{equation}\label{eq:e1}
L = -\frac{\text{d}E_{cool}}{\text{d}t}= \frac{64 \pi}{3} \frac{\sigma T_{rcb}^4 R_B^{\prime}}{\kappa \rho_{rcb}},
\end{equation}
where $\sigma$ is the Stefan-Boltzmann constant and $\kappa$ is the opacity at the $R_{rcb}$. \Cref{eq:E_cool,eq:e1} can be combined to yield a cooling timescale of the envelope, $t_{cool}$, given by $t_{cool}=|E_{cool}/(\text{d}E_{cool}/\text{d}t)|=E_{cool}/L$.
Following \citet{freedman2008a}, we model the opacity at the $R_{rcb}$ as $
\kappa/0.1\;\text{cm}^2\;\text{g}^{-1} = (\rho_{rcb}/10^{-3}\;\text{g}\;\text{cm}^{-3})^{0.6}$.

The energy required to lose the entire atmosphere is $E_{loss}\simeq gM_{atm}R_c$. Comparing this with the energy available for cooling given in \Cref{eq:E_cool} yields that $E_{cool} \lesssim E_{loss}$ for $M_{atm}/M_c > \mu/\mu_c \sim 5 \%$ (heavy atmospheres) and $E_{cool} \gtrsim E_{loss}$ for $M_{atm}/M_c < \mu/\mu_c \sim 5 \%$ (light atmospheres). Note, we ignored the $\gamma$ and $\gamma_c$ factors here for simplicity. This implies that, depending on the atmosphere to core mass-ratio after disk dispersal, planets can continue to evolve in two different ways. Planets with heavy atmospheres ($M_{atm}/M_c > \mu/\mu_c$) don't have enough energy to continually lose mass and their envelopes will cool and contract over time. In contrast, planets with light envelopes ($M_{atm}/M_c < \mu/\mu_c$) can, from an energy point of view, continue to lose mass over time. Furthermore, since for light envelopes $E_{cool} \gtrsim E_{loss}$ and since mass-loss proceeds at almost constant $\Delta R$ while decreasing the envelope density, atmospheric loss is a run-away process in the sense that energetically it gets easier with time (i.e., once the first half has been lost, it is even easier to loose the next half), ensuring that there is enough energy to lose the entire envelope \citep{ginzburg2016a,ginzburg2018a}.

However, despite sufficient energy, planets with light atmospheres are not necessarily stripped of their envelopes because, analogous to a Parker type wind, atmospheric mass-loss proceeds at a finite rate dictated by the escape rate of molecules at the Bondi radius \citep{ginzburg2016a,owen2016a}. Since the hydrodynamic flow needs to pass through the sonic point and since the mass flux is conserved, it is convenient to determine the mass-loss rate at the sonic point, $R_s=GM_p/2c_s^2$, where $c_s = (k_B T_{eq}/\mu)^{1/2}$ is the isothermal speed of sound. This yields a mass-loss rate of $\dot{M}=4 \pi \rho_s R_s^2 c_s$, where $\rho_s$ is the density at the sonic point, which can be related to the density at the radiative convective boundary by $\rho_s = \rho_{rcb} \text{exp} (-2R_s/R_{rcb})$ in the limit that $R_s>>R_{rcb}$. The mass-loss rate at the Bondi radius can therefore be written as 
\begin{equation}\label{eq:e2}
\dot{M}_{atm}^B = 4\pi R_s^2 c_s \rho_{rcb} \; \text{exp}\left( -\frac{GM
_p}{c_s^2 R_{rcb}}\right).
\end{equation}
From \Cref{eq:e2} we define the atmospheric mass-loss timescale as $t_{loss}=|M_{atm}/(\text{d}{M}_{atm}/\text{d}t)|$. This finite mass-loss rate is critical to the existence of planets in the valley and for explaining planets that have atmospheres of a few percent. The exponential dependence ensures that planets can hold on to their atmospheres because they did not have enough time for loss at larger orbital periods and/or that the cooling timescale can become shorter than the mass-loss timescale as a planet contracts during its evolution terminating any further mass-loss.

We follow the evolution of a given planet by simultaneously calculating its cooling and its atmospheric-loss due to core-powered mass-loss.
The energy of the planet decreases as a function of time as dictated by its internal luminosity, such that
\begin{align}
E_{cool}(t+\text{d}t) &=  E_{cool}(t) - L(t) \text{d}t,
\end{align}
where $L$ is given by \Cref{eq:e1}.
Similarly, the evolution of the atmospheric mass can be written as
\begin{align}
M_{atm}&(t+\text{d}t) = M_{atm}(t) - {min\left\lbrace\dot{M}_{atm}^E(t) , \dot{M}_{atm}^B(t) \right\rbrace} \text{d}t,
\end{align}
where the mass-loss rate at the Bondi radius is given by \Cref{eq:e2} and $\dot{M}_{atm}^E\simeq L(t)/g R_c$ is the energy-limited mass-loss rate. { The energy-limited mass-loss rate should be regarded as an absolute upper limit as it assumes that all of the cooling luminosity goes into driving the mass loss. In reality, the efficiency of this cannot be a 100\% since roughly half of the luminosity is radiated away, which is required, together with the irradiation from the star, to sustain the radiative-convective profile. The minimum of the energy-limited rate and the mass-loss rate at the Bondi radius} determines the rate at which atmospheric loss proceeds as it can be limited by the energy available for cooling or the escape rate of gas molecules from the Bondi radius.

\begin{table}
\centering
{\small
\begin{tabular}{c  c} 
\hline
Parameter &Value/Range (initial)\\ 
\hline
Orbital period ($P$) & [1, 100] days\\ 
Core radius ($R_c$) & [0.7, 4] $R_\oplus$\\ 
Core molecular mass ($\mu_c$) & 56 amu\\ 
Core adiabatic index ($\gamma_c$) & 4/3\\ 
Core characteristic density ($\rho_{c*}$) & $\rho_\oplus \sim$ 5.5 g cm$^{-3}$\\
Atmosphere molecular mass ($\mu$) & 2 amu\\ 
Atmosphere adiabatic index ($\gamma$) & 7/5\\ 
Host star & $\sim$ Sun\\
Integration time-step & $10^{-2}$ $\times$ min$\left\lbrace t_{cool}, t_{loss} \right\rbrace $ \\
Total evolution time & 3 Gyrs. \\
Number of planets & 1 million\\
\hline
\end{tabular}
\caption{Planet population and evolution parameters for our `reference' case.}
\label{table:sim_char}
}
\end{table}

\subsection{Modeling the Exoplanet Population}
Similar to previous works \citep{owen2017a,ginzburg2018a}, we adopt the following period and mass distribution when modeling the exoplanet population:
\begin{equation}{\label{eq:P_distr}}
\dv{N}{\;\text{log}P} \propto \begin{cases}
     P^{2}, & {P < 8\; \text{days}} \\
    \text{constant}, & {P > 8\; \text{days}}, \text{ and} 
  \end{cases}  
\end{equation}
\begin{equation}{\label{eq:M_c_distr}}
\dv{N}{M_c} \propto \begin{cases}
    M_c\;\text{exp} \left( -{M_c^2}/{(2 \sigma_{M}^2)} \right),&{M_c < 5\; M_\oplus}  \\
    M_c^{-2},&{M_c > 5\; M_\oplus}.
  \end{cases}  
\end{equation}
The planet mass distribution is described by a Rayleigh distribution with for planets less massive than $5 M_\oplus$ and as an inverse square tail for the planets more massive than $5 M_\oplus$. We use $\sigma_M=2.7 M_\oplus$ throughout this paper, unless stated otherwise. {We note  here that \citet{ginzburg2018a} investigated both $d{N}/d{M_c} = \text{[constant]}$ and $d{N}/d{M_c} = \text{[Rayleigh distribution]}$ for planets less massive than 5$M_\oplus$ and found no significant difference in the resulting 1-D radius distributions. For simplicity, we only investigate the latter in this work which is similar to the planet mass distribution used in published photoevaporation studies \citep[e.g.,][]{owen2017a}.}

{For the initial atmosphere to core mass-fraction ($f$) of the planets, we use 
\begin{equation}\label{eq:f}
f \simeq 0.05 (M_c/M_\oplus)^{1/2},
\end{equation}
which is motivated by a previous work on gas accretion and loss during disk dispersal \citep{ginzburg2016a}.}

The results presented in \Cref{sec:results} are based on the evolution of a population of a million planets over a period of 3 Gyrs. The defining parameters for our `reference' planet population and the numerical calculations for its evolution are summarized in \Cref{table:sim_char}. The planets in this `reference' population have rocky Earth-like cores, $H_2$ atmospheres and Sun-like host stars. Beyond this reference case, we explore a range of core compositions and planet-mass distributions. The choice of parameters for our reference case only differs from \citet{ginzburg2018a} in the explicit use of \Cref{eq:E_cool} for calculating $\Delta R$ instead of assuming the ratio of the core's and atmosphere's heat capacity to be {(17f)$^{-1}$}.

\section{Results} \label{sec:results}
In this section, we present the results of our evolution model described in \Cref{sec:model}. First we discuss the results for the `reference' population as defined in \Cref{table:sim_char} and then investigate the dependence of our findings on core compositions and planet-mass distributions.

\begin{figure*}
\centering
\includegraphics[width=0.30\textwidth,trim=360 245 840 620,clip]{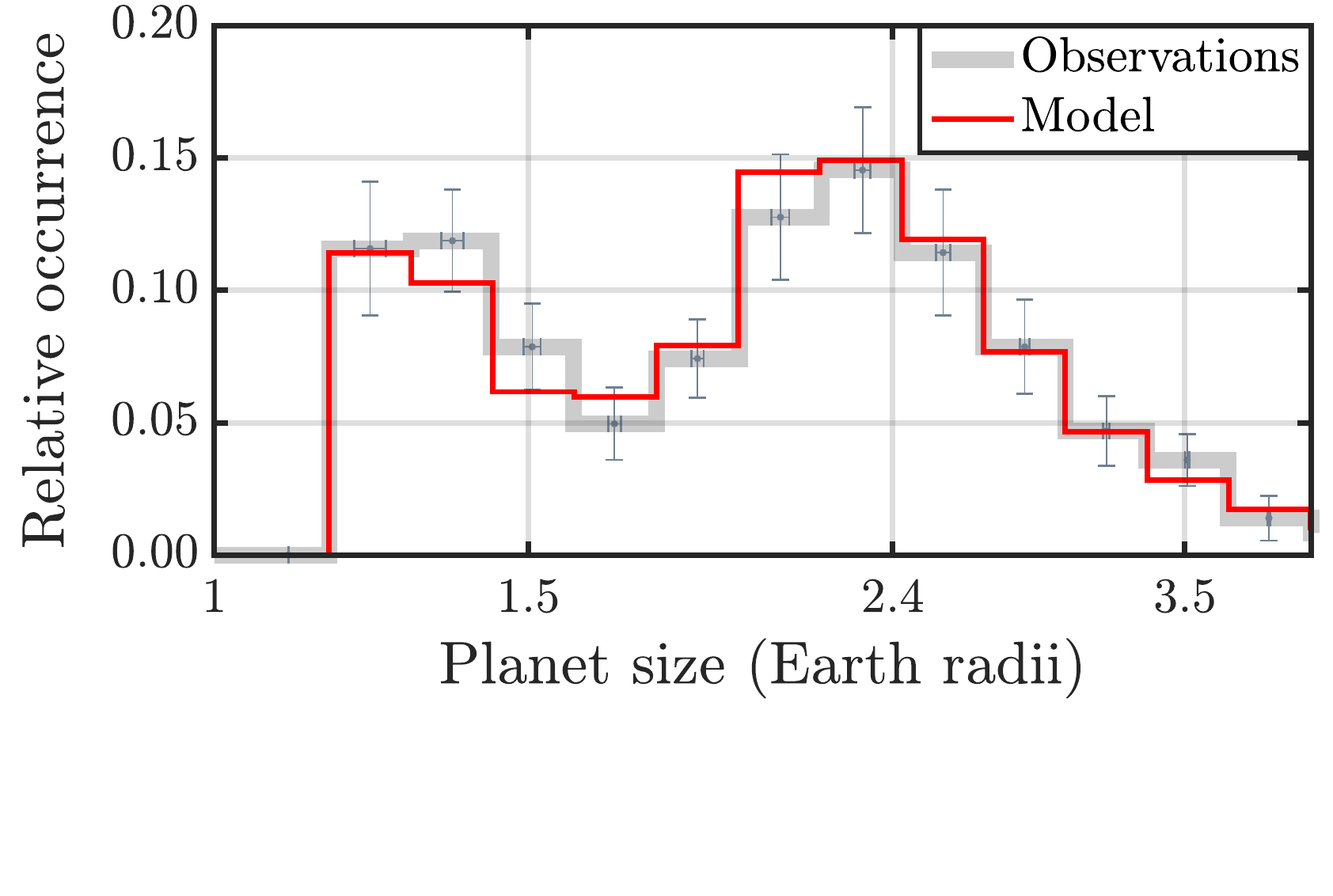}
\includegraphics[width=0.35\textwidth,trim=230 385 860 400,clip]{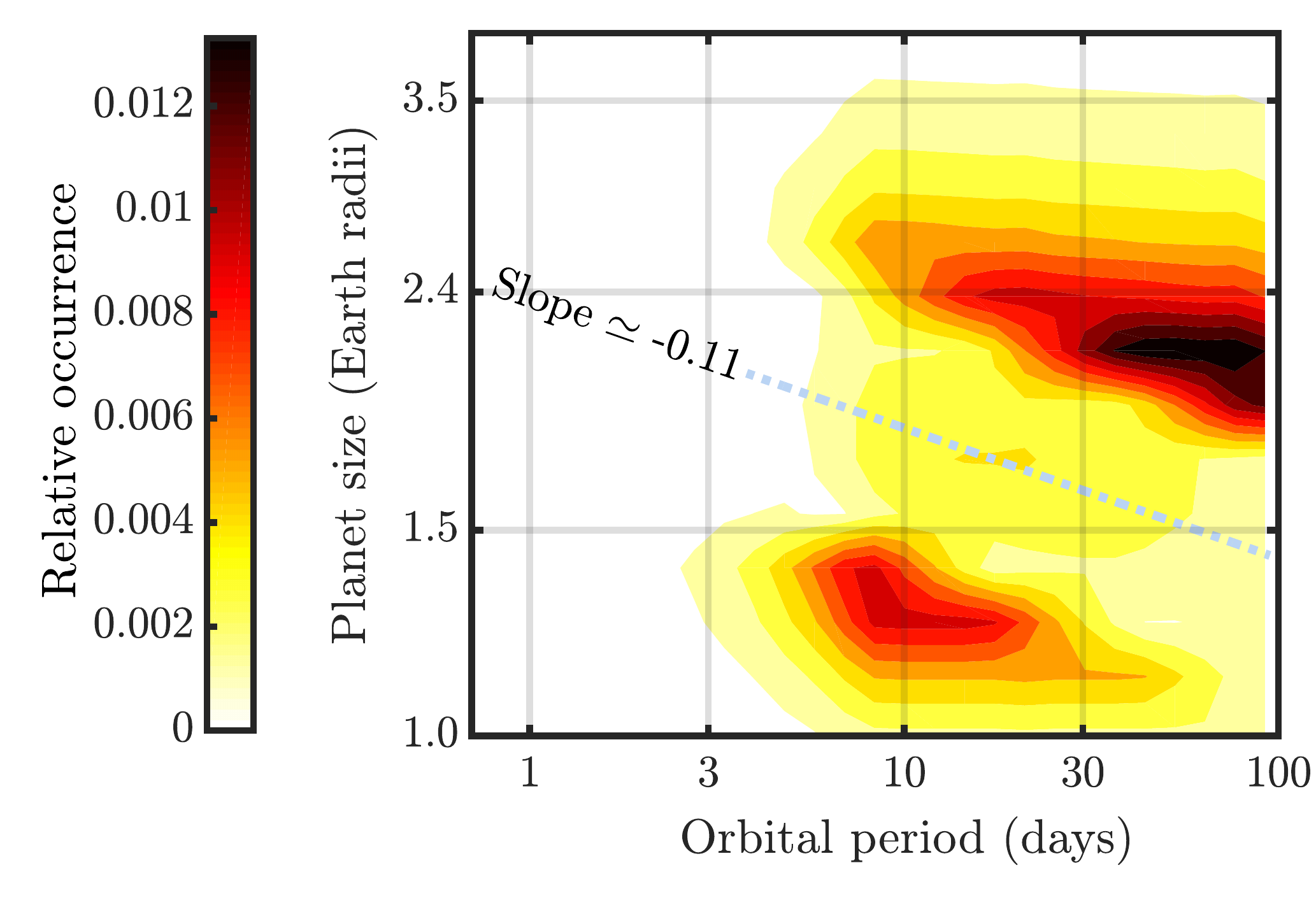}
{ \includegraphics[width=0.32\textwidth,trim=60 552 310 60,clip]{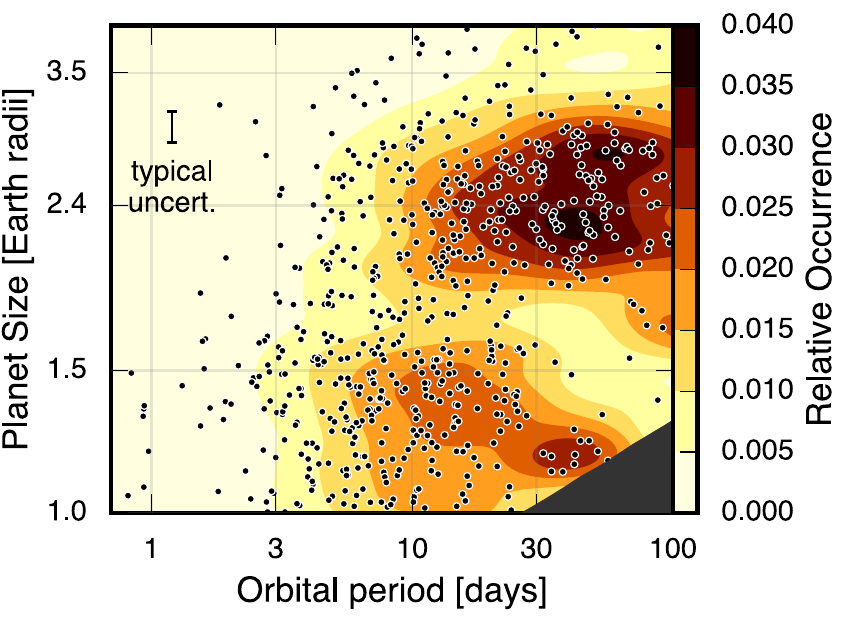}}
\caption{Comparing core-powered mass-loss results with observations. \textit{Left panel}: Histogram of planet size (radii). The two histograms correspond to the results from our evolution model for our `reference' planet population (see Table 1) shown  in red and observations shown in gray \citep[see Table 3]{fulton2017a}. \textit{Middle and right panels}: Two-dimensional distributions of planet size and orbital period. The middle panel displays the results from our core-powered mass-loss evolution model, while the right panel corresponds to observations \citep[from][with permission]{fulton2018a}. The dashed blue line in the middle panel corresponds to the center of the valley. Its slope is given by  $\text{d log} R_p/ \text{d log} P \simeq -0.11$. The results from our core-powered mass-loss model are in good agreement with the observations.}
\label{fig:obs_comparison}
\end{figure*}

\subsection{Comparing the Core-powered Mass-loss Results with Observations} \label{sec:compare_obs}

\citet{ginzburg2018a} already demonstrated that the core-powered mass-loss mechanism itself produces a bimodal exoplanet radius distribution and that it yields results consistent with the observed valley in the radius distribution of close-in super-Earths and sub-Neptunes. Here we extend this comparison from a single histogram of planet radii to a two-dimensional orbital period-planet radius parameter space and use our results to infer properties of the observed exoplanet population.

\Cref{fig:obs_comparison} displays our core-powered mass-loss results and compares it with the observations from \citet{fulton2017a} and \citet{fulton2018a}. The left panel shows the histogram of relative occurrence of planet radii from our model (red) and observations (grey) \citep{fulton2017a}. To facilitate the comparison between our results and observations, we display and normalize our results over the same planet radius range as for the observations shown in \Cref{fig:obs_comparison}. As shown in previous work \citep{ginzburg2018a}, we find good agreement between the radius distribution produced by the core-powered mass-loss mechanism and the observed exoplanet population. The left panel of \Cref{fig:obs_comparison} shows that the valley is located between $\sim$ 1.5-2.0 $R_\oplus$ with a width of $\sim$ 0.5 $R_\oplus$. The lower peak of the `super-Earths', i.e., planets stripped of their envelopes, is at 1.2-1.4 $R_\oplus$ and the higher peak of the `sub-Neptunes', i.e., planets that kept most of their atmospheres, is at 2.0-2.7 $R_\oplus$. 

The middle panel of \Cref{fig:obs_comparison} presents our results in the two-dimensional parameter space of planet size and orbital period. The right panel shows the observational results from \citet{fulton2018a} for the same two-dimensional parameter space. We generally find good agreement with observations. Specifically, our results display a valley of approximately constant width that moves to smaller planet radii with increasing orbital period. This is a manifestation of the decreasing susceptibility of planets to lose their atmospheres with increasing orbital period. This results in a negative slope for the valley, which is plotted as a dashed line in the middle panel. We find a slope $\text{d log} R_p/ \text{d log} P \simeq -0.11$ both analytically and numerically, which is in excellent agreement with observations reported by \citet{eylen2018a} { and \citet{MCG19} who find $\text{d log} R_p/ \text{d log} P = -0.09^{+0.02}_{-0.04}$ and $\text{d log} R_p/ \text{d log} P = -0.11^{+0.03}_{-0.03}$, respectively.} We discuss the physical processes determining the slope and analytically derive the slope of the valley in \Cref{sec:slope_valley}.

The degree of similarity between observations and the core-powered mass-loss results presented here demonstrates that the core-powered mass-loss mechanism can by itself reproduce the observed valley in the exoplanet radius distribution and this is not unique to photoevaporation \citep{owen2017a}. Specifically, \Cref{fig:obs_comparison} shows that the core-powered mass-loss mechanism can reproduce the valley in radius-period space, its position, shape and slope, and the location and magnitude of the peaks of the exoplanet population on either side of the valley.

\subsection{Slope of the Valley} \label{sec:slope_valley}
As discussed in \Cref{sec:CPM}, for a planet to lose its envelope it not only has to have enough energy to unbind the atmosphere but it also needs to have enough time for the mass-loss to occur.

{ \subsubsection{Mass-loss limited by time, $t_{cool}=t_{loss}$, $P \gtrsim 8$ days}
 We find that it is the time-limit and not the energy-limit that dictates which planets lose and retain their atmospheres for orbital periods of about 8 days and larger. It is, to first order, the mass-loss timescale, due to its exponential dependence on period and planet size, that creates the valley and determines its slope in the planet radius and orbital period space.} Specifically, we find that the criterion $t_{cool}=t_{loss}$ separates the planets that will end up above and below the valley.
 
 { Planets accrete their envelopes with $R_{rcb} \sim R_{B}$ \citep[e.g.][]{lee2015a,ginzburg2016a} and then shed their outer layers and shrink in radius during and after the disk dispersal phase \citep[e.g.][]{owen2016a,ginzburg2016a}. This ensures that, from a planet formation point of view, initially all planets start out with $t_{loss} < t_{cool}$, i.e. initially the envelope mass-loss timescale is short because $R_{rcb}\sim R_{B}$ and hence the exponent in the mass-loss timescale is small. However, as planets initially lose mass, they also shrink, increasing the mass-loss timescale, and in some cases the cooling timescale catches up with the mass loss timescale such that $t_{cool}=t_{loss}$. The criterion $t_{cool}=t_{loss}$ separates the planets that will end up above and below the valley  because once planets can cool faster than they can lose mass (i.e., $t_{cool} < t_{loss}$), they shrink in size and any subsequent mass-loss is hence cut-off rapidly due to the exponential dependence on planet size of the mass-loss timescale (see \Cref{sec:model})}. Furthermore, due to the exponential dependence of the mass-loss timescale, it is, to first order, the exponent that determines the slope of the valley in the radius-period parameter space. Setting the mass-loss timescale equal to the cooling timescale, we have from \Cref{eq:e2} that $GM_p/c_s^2R_{rcb} \simeq \rm{constant}$ and hence
\begin{equation}\label{eq:slope_approx}
\frac{R_c^4 P^{1/3}}{R_p} \rho_{c*}^{4/3}\simeq {R_p^3 P^{1/3}} \rho_{c*}^{4/3} = \text{constant},
\end{equation}
where we substituted for the speed of sound and the mass-radius relation of the core and used the fact that $R_{rcb} = R_{p}  \simeq 2 R_c$. As long as a planet's core density has no semi-major axis dependence, \Cref{eq:slope_approx} yields $R_p\propto P^{-1/9}$ which corresponds to a slope in the logarithmic parameter space of planet radius and orbital period of
\begin{equation}
\frac{\text{d log}R_p}{\text{d log}P} = -\frac{1}{9} \simeq -0.11.
\end{equation}
This is in excellent agreement with the observed slope reported by \citet{eylen2018a}, $\text{d log} R_p/ \text{d log} P = -0.09^{+0.02}_{-0.04}$, based on high precision asteroseismic measurements of stellar parameters { and results obtained by \citet{MCG19} who report a slope of $\text{d log} R_p/ \text{d log} P = -0.11^{+0.03}_{-0.03}$}. The middle panel of \Cref{fig:obs_comparison} shows a dashed line denoting the center of the observed valley. The slope of this line, as measured from our numerical simulations, is in full agreement with our analytical estimate above. \Cref{eq:slope_approx} shows that the valley's slope in the radius-period space does, to first order, not depend on $f$, properties of the host star or the core density as these only change the constant in \Cref{eq:slope_approx} but not the power-law relation between $R_p$ and $P$. These quantities, do however, change the location of the valley, which is set by the constant in \Cref{eq:slope_approx} and which we will come back to when examining the dependence of our results on the core composition in \Cref{sec:core_composition}.

\begin{figure*}
\centering
\includegraphics[width=\textwidth,trim=250 450 300 370,clip]{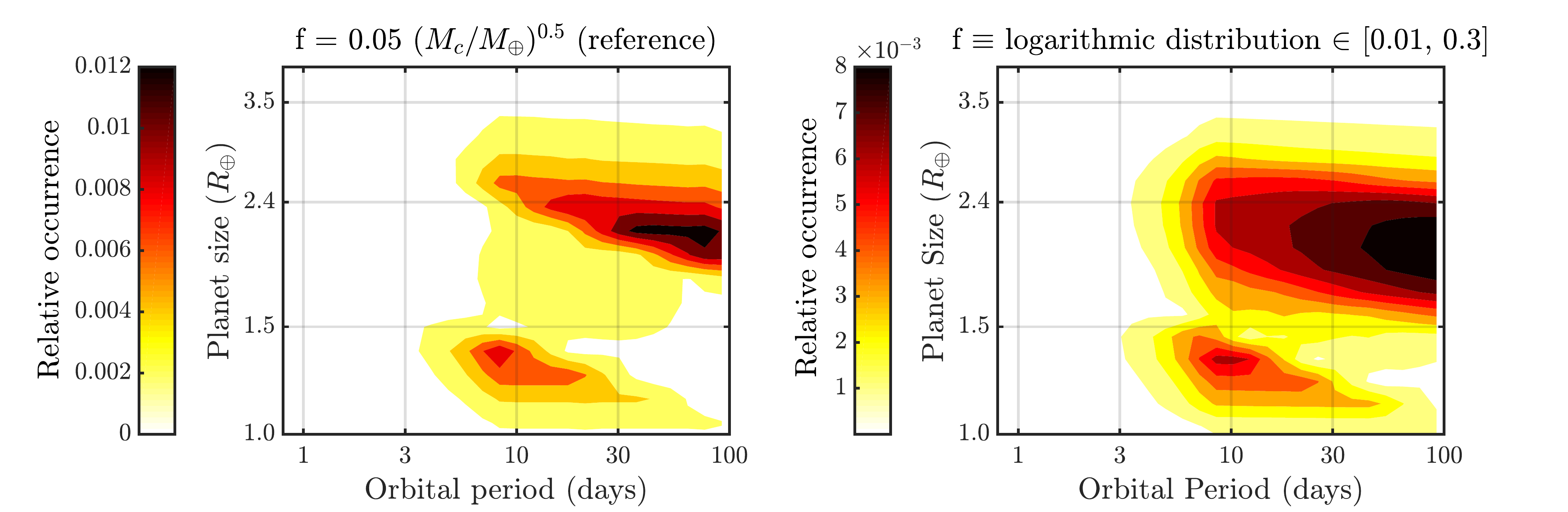} 
\includegraphics[width=\textwidth,trim=250 450 300 370,clip]{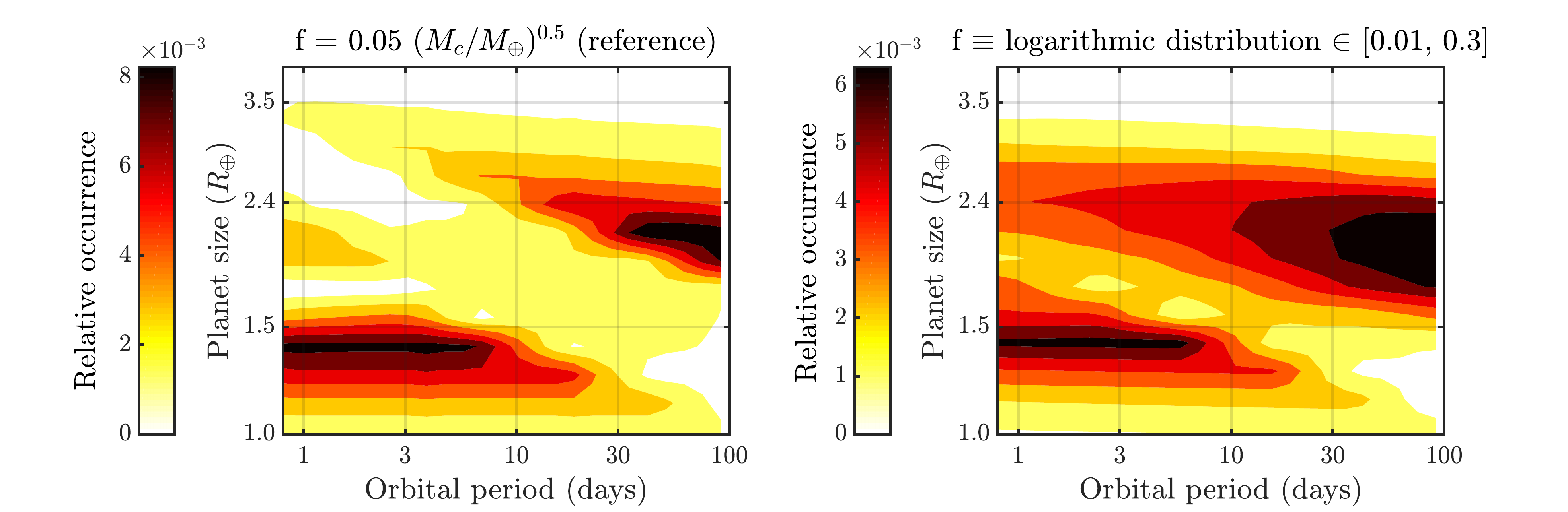} 
\caption{Core powered mass-loss results for different distributions of initial envelope fractions, $f$. The left column corresponds to envelope fractions distributed according to Equation (\ref{eq:f}) and the right column to the logarithmic distribution used in \citet{owen2017a}. The top panel shows the results weighted by the exoplanet period distribution given by Equation (\ref{eq:P_distr}) and the bottom panel shows, for clarity, the results for a uniform period distribution. For the $f$ distribution used in this work (see Equation (\ref{eq:f})), the transition between the energy-limited and the time-limited mass-loss regime occurs around 8 days (see flat lower edge of the valley for periods of less than about 8 days in bottom left panel of the figure). If there is no unique relationship between a planet's core mass and its envelope fraction (e.g. in the logarithmic distribution for $f$ used in the right panel), then there is no flat line that sets the lower edge of the radius valley at short orbital periods (see bottom right side of the panel) and there is no single orbital period that marks the transition between the energy-limited and time-limited mass loss regime (see Section \ref{sec:slope_lt_8} for details).}
\label{fig:f_distr}
\end{figure*}

{ It may at first seem surprising that the slope that is obtained by setting $t_{cool}=t_{loss}$ is not significantly modified over time as the planets continue to thermally evolve and contract. However, we find both analytically and numerically that the contraction rate, for planets that satisfy $t_{cool}=t_{loss}$, only weakly depends on $R_c$ and $f$ for the parameters investigated here. Furthermore, even if a logarithmic distribution of $f$ is assumed, the slope of the valley is, to first order, still well described by setting $t_{cool}=t_{loss}$ as is illustrated by the results shown in Figure \ref{fig:f_distr}. The logarithmic $f$ distribution does change the shape of the valley somewhat, especially the upper edge which is due to the fact that, for a logarithmic $f$ distribution, the evolution rates start to differ significantly for a given planet mass.} 

The one quantity that does change the slope of the valley is the mass-radius relationship of the core. Specifically, we find from \Cref{eq:e2,eq:slope_approx} that
\begin{equation}
\frac{\text{d log}R_p}{\text{d log}P} \simeq \frac{1}{3(1-\beta)}, \text{ where } M_c \propto R_c^{\beta}.   
\end{equation}
Precise observational measurements of the valley's slope are therefore able to determine the exoplanet mass-radius relation of the core. We find that published measurements of the slope \citep{eylen2018a} are in agreement with $M_c/M_{\oplus} \propto (R_c/R_{\oplus})^4$ but inconsistent with $M_c/M_{\oplus} \propto (R_c/R_{\oplus})^3$, highlighting the significance of internal compression of massive cores.

{ \subsubsection{Mass-loss limited by energy, $E_{cool}=E_{loss}$, $P\lesssim 8$days}\label{sec:slope_lt_8}

As shown in \Cref{sec:CPM}, whether a planet has enough energy to lose its envelope is solely dictated by the envelope to core mass-fraction, $f$. If $f$ has, to first order, no dependence on the distance from the host star, as we assume in \Cref{eq:f}, then the maximum envelope fraction for which significant atmospheric loss can occur corresponds to a single planet mass independent of period provided that there is a unique relationship between $f$ and the core mass (as, for example, given in \Cref{eq:f} and predicted by planet formation models of atmospheric accretion of super-Earths and sub-Neptunes \citep{lee2015a,ginzburg2016a}). Evaluating $E_{loss}= E_{cool}$ and accounting for all the $\gamma$ dependencies yields $M_{atm}/M_c \simeq 14\%$ for $\gamma=7/5$. This implies that planets with $f \lesssim 14\%$ have enough energy available from cooling that they can unbind their H/He envelopes. This envelope fraction can be converted into a core mass and radius using \Cref{eq:f} which yields $M_c\simeq8 M_{\oplus}$ and $R_c\simeq1.7 R_{\oplus}$, respectively. We therefore expect a single flat line for the lower edge of the valley corresponding to $R_c\simeq1.7 R_{\oplus}$ for planets whose atmospheric loss is energy and not time-limited, which corresponds to planets with orbital periods of less than about 8 days  (see the left panel in \Cref{fig:f_distr}). 

We note here, that if there is no unique relationship between a planet's core mass and its envelope fraction, like, for example, in the logarithmic distribution for $f$ used in \citet{owen2017a}, then the critical atmosphere to core mass ratio, for which significant atmospheric loss will occur in the energy-limited regime, cannot be converted to a single core mass and radius. Hence, in this case, there is no flat line that sets the lower edge of the radius valley and there is no single orbital period at which the transition from energy-limited to time-limited atmospheric loss occurs (see the right panel in \Cref{fig:f_distr}). 

For the parameters used in this work, the transition between the energy-limited and time-limited mass-loss occurs around 8 days, but the exact transition in period space depends on the relationship between $f$ and the core mass (see \Cref{fig:f_distr} and \Cref{eq:f}). This implies that if the lower edge of the radius valley is observed to be a flat line for planets inside orbital periods of several days than this can be used to constrain the initial envelope fractions from planet formation and hence the atmospheric accretion process itself.}

\begin{figure*}
\centering
\includegraphics[width=\textwidth,trim=210 320 190 310,clip]{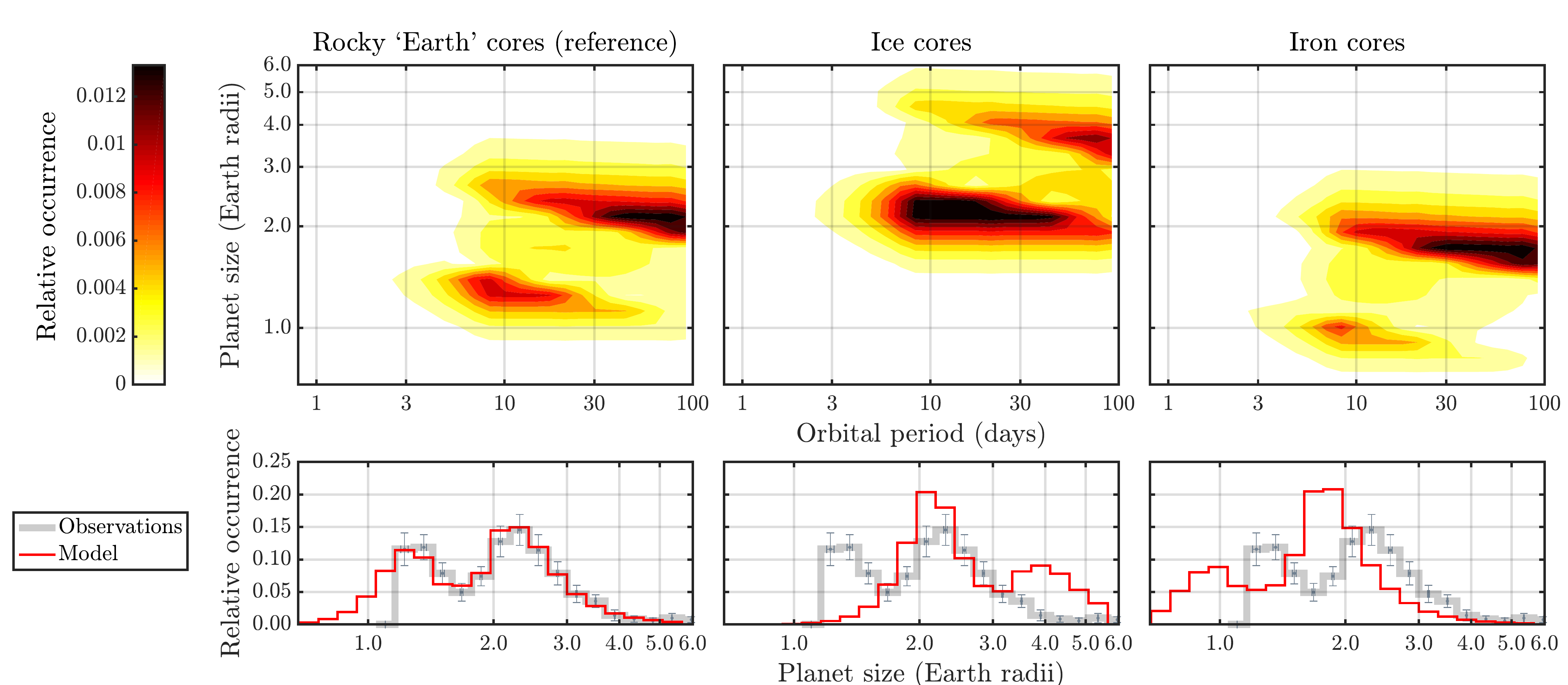} 
\caption{Dependence of the core-powered mass-loss results on core composition. This figure displays two-dimensional distributions of planet size and orbital period in the top row, and histograms of planet size in the bottom row. The three columns correspond to three different core materials, namely (from left to right), rocky `Earth' like (reference case; $\mu_c=56\;amu,\;\rho_{c*}=5.5\;\text{g cm}^{-3}$), ice ($\mu_c=18\;amu,\;\rho_{c*}=1.3\;\text{g cm}^{-3}$) and iron ($\mu_c=56\;amu,\;\rho_{c*}=11\;\text{g cm}^{-3}$). While the case with rocky cores closely resembles the observations (see \Cref{fig:obs_comparison}), for icy and iron cores the valley shifts to higher and lower planet sizes, respectively. To aid the comparison between our results and observations, we normalized our findings over the same planet radius range as the observations, but display our numerical results down to planet sizes that are smaller than the smallest observed radius bin in \citet{fulton2017a}.}
\label{fig:core_material}
\end{figure*}

\subsection{Constraints on the Core Composition}\label{sec:core_composition}

We investigate the dependence of our results on composition of the exoplanet cores. We account for changes in core material by modifying the core density ($\rho_{c*}$) and its molecular mass ($\mu_c$). 

\subsubsection{Single Composition Cores}
As we can see from \Cref{eq:slope_approx}, the slope of the valley should be insensitive to changes in core density, $\rho_{c*}$, but its location should scale as $\rho_{c*}^{-4/9}$, or more generally as $\rho_{c*}^{-\beta/9}$, where $\beta$ is the exponent in the mass-radius relation of the core. We therefore expect the slope to remain the same but the location of the valley to move to larger planet radii for lower density cores and to smaller planet radii for high density cores. This is exactly what we find numerically in our results shown in \Cref{fig:core_material} for cores made of ice and iron. From the scaling with core composition above, we predict that the bottom of the valley should move up by a factor of 1.9 from about $1.6 R_p$ to about $3.0 R_p$ for icy core and move down by a factor of 0.7 for iron cores from about $1.6 R_p$ to about $1.1 R_p$ compared to our rocky `reference' case, which is indeed what is shown in \Cref{fig:core_material}. Another implication of the exponential dependence on core density is that the characteristics of the final radius distribution are, to first order, not affected by the changes in $\mu_c$.

Finally, our numerical results in \Cref{fig:core_material} confirm that, while there are significant changes in the location of the valley for the different core compositions, the slope of the valley remains essentially unchanged, as expected.

\begin{figure}
\centering
\includegraphics[width=.45\textwidth,trim=220 320 880 300,clip]{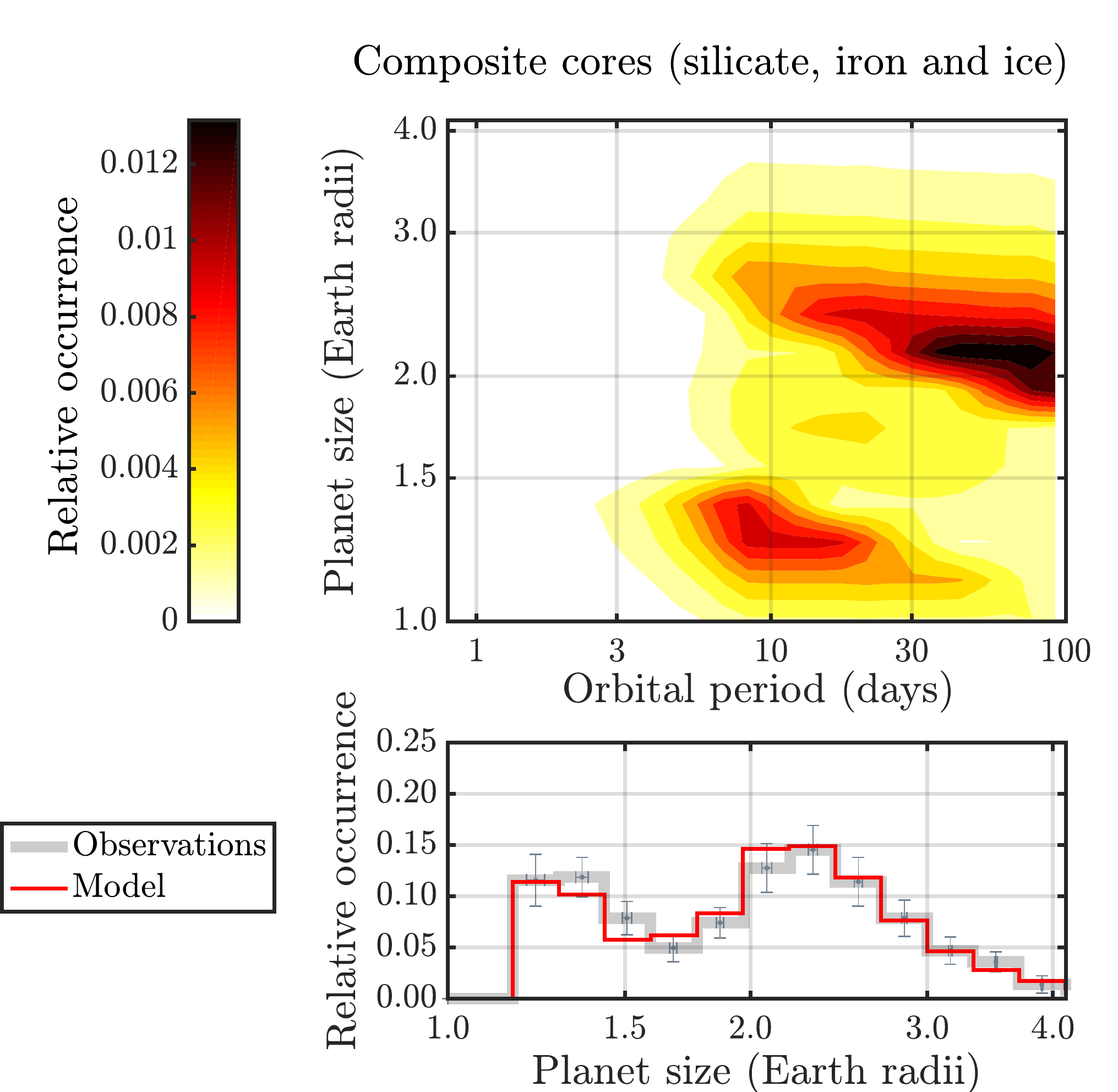} 
\caption{Maximum water-ice content for Earth-like cores. This figure shows a two-dimensional distribution of planet size and orbital period in the top panel and histogram of planet size in the bottom panel. The cores of the planet population have the following composition: 56\% silicate ($\mu_c=76$ amu, $\rho_{c*}=4\;\text{g cm}^{-3}$), 28\% iron ($\mu_c=56$ amu, $\rho_{c*}=11\;\text{g cm}^{-3}$) and 16\% ice ($\mu_c=18$ amu, $\rho_{c*}=1.3\;\text{g cm}^{-3}$) with an effective $\mu_c=61$ amu and $\rho_{c*}\simeq5.5\;\text{g cm}^{-3}$. Earth-like composition cores can contain up to $\sim 20\%$ of water-ice and still match the observations.}
\label{fig:mimicking_obs}
\end{figure}

\subsubsection{Maximum Water/Ice Content of Super-Earths and Sub-Neptune Cores} 
The strong dependence of the valley's location on the density of planetary cores implies that we can, not only constrain the bulk composition of the cores of super-Earths and sub-Neptunes, but that we can also place limits on their maximum water/ice content. We demonstrate this in an example in which we assume a core composition that initially consists of $1/3$ iron ($\rho_{c*} = 11$ g cm$^{-3}$)  and $2/3$ silicate ($\rho_{c*} = 4$ g cm$^{-3}$) by mass, and then add the maximum amount of water-ice ($\rho_{c*} = 1.3$  g cm$^{-3}$) that can reproduce the observations. \Cref{fig:mimicking_obs} shows that $ 16\%$ of water-ice can be added to Earth-like composition cores without causing a noticeable discrepancy between our core-powered mass-loss results and the observations. This implies that, first, the location of the valley constrains the bulk density of the super-Earth and sub-Neptune population and, second, this in turn can be used to place limits on their possible compositions. Overall we find, similar to photoevaporation studies \citep[e.g.,][]{owen2017a,JM18}, that cores must be predominantly rocky with water-ice fractions of less than $\sim 20\%$.

In addition, we can conclude from the location of the valley and the peaks for the different core compositions in \Cref{fig:core_material} that the fraction of pure water/ice worlds and pure iron cores must be relatively small. 

These inferences imply that most of the close-in super-Earths and sub-Neptunes formed in a water/ice poor environment.

\begin{figure*}
\centering
\includegraphics[width=\textwidth,trim=210 320 190 300,clip]{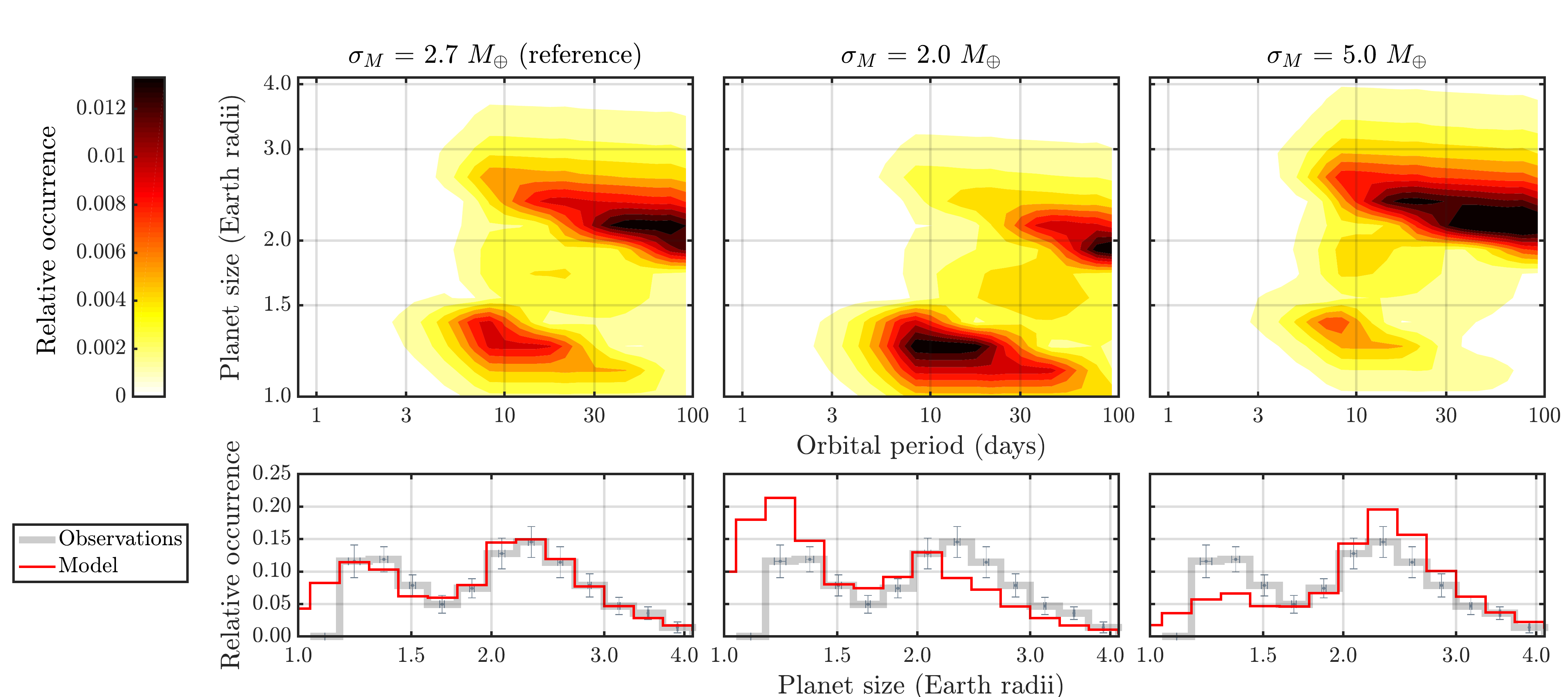} 
\caption{Dependence of the core-powered mass-loss results on planet-mass distribution. Figure shows two-dimensional distributions of planet size and orbital period in the top row, and histograms of planet size in the bottom row. The three columns correspond to three different planet-mass distributions modeled as a Rayleigh distribution with an inverse-square tail, with $\sigma_M$ values of 2.7 M$_\oplus$ (reference case, left panel), 2.0 M$_\oplus$ (middle panel) and 5.0 M$_\oplus$ (right panel), see \Cref{eq:M_c_distr} for details. As, expected, for a lower $\sigma_M$, the occurrence of planets below the valley is larger than in the `reference' case. In contrast, the peak above the valley is more pronounced for the higher $\sigma_M$ value than in the `reference' case. To aid the comparison between our results and observations, we normalized our findings over the same planet radius range as the observations, but display our numerical results down to planet sizes that are smaller than the smallest observed radius bin in \citet{fulton2017a}.}
\label{fig:core_distr}
\end{figure*}

\subsection{Dependence on Planet-mass Distribution}

We also investigate the sensitivity of our results to the underlying distribution of planet masses ($M_p\simeq M_c$). Specifically, we keep the Rayleigh distribution and the inverse-square tail, but change the value of $\sigma_M$; see \Cref{eq:M_c_distr}. \Cref{fig:core_distr} displays our `reference' case (left panel) and results for $\sigma_M=2.0 M_\oplus$ and $\sigma_M=5.0 M_\oplus$ in the middle and right panel, respectively. \Cref{fig:core_distr} shows that for a lower $\sigma_M$, i.e., for an underlying distribution peaking at a lower planet mass, the peak below the valley is more significant than in the `reference' case. In contrast, for higher values of $\sigma_M$ the peak above the valley is more pronounced. However, any changes in the mass distribution do not fundamentally change the location of the valley itself. This implies that the location of the valley does not depend on the detailed assumptions of the planet-mass distribution (as long as it is chosen to cover the observed parameter space in planet radii/masses), but is determined by the planet's core composition instead (see \Cref{fig:core_material} and discussion in \Cref{sec:core_composition}). In contrast, the relative magnitude of the peaks above and below the valley is sensitive to the details of the underlying planet-mass distribution and it can hence be used to constrain the planet population from observations. Similar to previous studies, we find a Rayleigh distribution with an inverse-square tail, with $\sigma_M \sim 3.0 M_\oplus$, yields a good fit to the observations; see \Cref{fig:obs_comparison} and \Cref{fig:core_distr}.

\section{Discussion \& Conclusions}\label{sec:conclusion}
Close-in exoplanets display an intriguing gap in their radius distribution around 1.5-2.0 Earth radii \citep{owen2013a,fulton2017a,fulton2018a,eylen2018a}. 

In this work, we numerically followed the thermal evolution and atmospheric loss of small, short-period planets modeled on the observed exoplanet population. To distinguish our results from any atmospheric loss due to photoevaporation, we only considered the planet's evolution due to its own cooling luminosity and its subsequent mass-loss, i.e., we focus on the planet's evolution due to the core-powered mass-loss mechanism \citep{ginzburg2016a,ginzburg2018a}.

We demonstrated that planetary evolution based on the core-powered mass-loss mechanism alone (i.e., without any photoevaporation) is capable of reproducing the observed valley in the radius distribution of small, close-in planets. In particular, we are able to match the valley's position, shape, slope and the relative magnitude of the peaks above and below the valley. Our results are in good agreement with observations both when examining the histogram of planet sizes and the two-dimensional planet size-orbital period parameter space. Our findings imply that even super-Earths that appear as barren rocky cores today likely formed with primordial hydrogen and helium envelopes and that they are therefore not true terrestrial planet analogs from the point of view of their formation. 

We analytically derive the slope of the valley by equating the atmospheric mass-loss timescale to the cooling timescale and find a slope for the valley $\text{d log} R_p/ \text{d log} P \simeq -0.11$. This is identical to the slope that we find from our numerical evolution models and is in good agreement with the value reported by \citet{eylen2018a}, $-0.09^{+0.02}_{-0.04}$ { and \citet{MCG19}, $-0.11^{+0.03}_{-0.03}$}.

We find, both numerically and analytically, that the slope of the valley is, to first order, independent of the core density and planet-mass distribution, but that it does depend on the mass-radius relation of the core. Precise observational measurements of the valley's slope should therefore probe the exoplanet mass-radius relation of the core. We find that published measurements of the slope are in agreement with $M_c/M_{\oplus} \propto (R_c/R_{\oplus})^4$ but inconsistent with $M_c/M_{\oplus} \propto (R_c/R_{\oplus})^3$, highlighting the significance of internal compression of massive cores.

In addition to understanding the formation of the valley itself, we investigated the dependence of our results on core composition and planet-mass distribution, and compared our findings with observations from recent exoplanet studies \citep{fulton2017a,fulton2018a,eylen2018a}.

By varying the density and mean molecular mass of the cores, we demonstrated analytically and numerically that the location of the valley depends primarily on the core's density and that it shifts to larger (smaller) planetary radii for lower (higher) density cores. This implies that the location of the valley constrains, to first order, the bulk density of the cores of the super-Earth and sub-Neptune population. We find that cores must be predominantly rocky with typical water-ice fractions of less than $\sim 20\%$ to match observations. In addition, we conclude from the location of the valley and the peaks for the different core compositions that the fraction of water worlds and iron cores must be relatively small. These inferences imply that most of the close-in super-Earths and sub-Neptunes accreted predominantly inside the ice line.

We also investigated the sensitivity of our results to the underlying distribution of planet masses and discovered that the location of the valley does not depend on the detailed assumptions of the planet-mass distribution (as long as it is chosen to cover the observed parameter space in planet radii/masses). In contrast, the relative magnitude of the peaks above and below the valley is sensitive to the details of the planet-mass distribution and it can hence be used to constrain the planet population from observations. Similar to previous studies, we find that the planet-mass distribution modeled as a Rayleigh distribution with an inverse-square tail, with $\sigma_M \sim 3.0 M_\oplus$, can closely reproduce observations \citep[e.g.][]{fulton2017a,fulton2018a,eylen2018a}.

{ Throughout this work, we use $\gamma=7/5$ which corresponds to molecular gas. However, temperatures deep inside the envelope, especially early on, can be hot enough for hydrogen to exist as monatomic gas for which $\gamma=5/3$ and the dissociation of hydrogen in the atmosphere even allows for the possibility of values of $\gamma < 4/3$. Although we chose to model our atmospheres with $\gamma=7/5$ the results presented in this work are general and do not depend on the exact value of $\gamma$ used. The main way in which the exact choice of $\gamma$ matters is that it determines how the mass and energy is distributed in the atmosphere. However, because we are investigating planets after the spontaneous mass-loss/boil-off phase, the radius of the atmosphere, given by $R_{rcb}$, varies only between one and and a few $R_c$ which implies that the actual variations between the monatomic and diatomic case for the mass distribution in the atmosphere are rather small. In addition, since the exponent given in Equation \ref{eq:slope_approx} that determines the slope of the valley does not depend on $\gamma$, the slope of the radius valley is the same regardless of the value of $\gamma$ used. Possibly, the most interesting difference between the monatomic and diatomic case relevant for this work is the envelope-to-core-mass ratio for which $E_{loss}\simeq E_{cool}$ (see Sections \ref{sec:planet_structure} and \ref{sec:CPM} for details). However, even in this case the difference is small. Evaluating $E_{loss}= E_{cool}$ accounting for all the $\gamma$ dependencies yields $M_{atm}/M_c \simeq 14\%$ and $M_{atm}/M_c\simeq15\%$, for $\gamma=7/5$ and $\gamma=5/3$, respectively. These envelope fractions can be converted into core radii using Equation \ref{eq:M_atm} which yields $R_c\simeq1.7 R_{\earth}$ for both cases. Finally, the evolution timescales are longer by a factor of several for $\gamma=5/3$. However, this longer evolution timescale would have the most significant effect on Gyr timescales by which time most planets will have cooled sufficiently such that the $\gamma=7/5$ case investigated here likely provides a better description of their envelopes than the  $\gamma=5/3$.}

In this study, we demonstrate that a planet's own cooling luminosity is capable of reproducing the observed valley in the radius distribution of close-in planets. Although atmospheric loss by the core-powered mass-loss mechanism seems an inevitable by-product of atmospheric accretion and the planet formation process itself, our results should not be taken to imply that atmospheric loss by photoevaporation does not happen or that it has to be unimportant. In fact, it seems likely that both processes contributed to sculpting the observed exoplanet population. Our work demonstrates that the core-powered mass-loss mechanism \citep{ginzburg2016a,ginzburg2018a} yields similar results to the photoevaporation mechanism \citep{owen2017a} in terms of the existence, location and slope of the radius valley, core composition, and the core mass distribution. This implies that, regardless of which of the two mechanism dominates (if any), the conclusions concerning, for example, the core composition and the implications that most super-Earths and sub-Neptunes are water-ice poor, are independent of the mass-loss mechanism.

In future work, we plan to combine photoevaporation with the core-powered mass-loss mechanism, and to extend the current investigation to a range of stellar types with the hope to be able to disentangle the specific signatures that these two mechanisms leave in the observed exoplanet population. {This work will also address whether the core-powered mass-loss mechanism supports claims from photo-evaporation models that the planet-mass distribution varies with stellar mass \citep{wu2018a} and it will examine the radius valley's dependence on metallicity \citep{owen2018a}.}

\section{Acknowledgement}
\par { We thank James Owen for valuable comments that helped to improve the manuscript.}
H.E.S. thanks Sivan Ginzburg and Re'em Sari for insightful discussions and gratefully acknowledges support from the National Aeronautics and Space Administration under grant No. $17~\rm{XRP}17\_~2-0055$ issued through the Exoplanet Research Program.

\bibliographystyle{mnras}
\bibliography{planet_evo}

\bsp	
\label{lastpage}
\end{document}